# POLYOMINO BASED DIGITAL HALFTONING

David VANDERHAEGHE
*ARTIS - INRIA - Grenoble Universities*
david.vanderhaeghe@inrialpes.fr

Victor OSTROMOUKHOV
*Université de Montréal*
ostrom@iro.umontreal.ca

**ABSTRACT**

In this work, we present a new method for generating a threshold structure. This kind of structure can be advantageously used in various halftoning algorithms such as clustered-dot or dispersed-dot dithering, error diffusion with threshold modulation, etc. The proposed method is based on rectifiable polyominoes -- a non-periodic hierarchical structure, which tiles the Euclidean plane with no gaps. Each polyomino contains a fixed number of discrete threshold values. Thanks to its inherent non-periodic nature combined with off-line optimization of threshold values, our polyomino-based threshold structure shows blue-noise spectral properties. The halftone images produced with this threshold structure have high visual quality. Although the proposed method is general, and can be applied on any polyomino tiling, we consider one particular case: tiling with G-hexominoes. We compare our polyomino-based threshold structure with the best known state-of-the-art methods for generation threshold matrices, and conclude considerable improvement achieved with our method.

## 1. INTRODUCTION

Digital halftoning is a well-established technique for visualization of continuous tone or rich multiple-tone images on visualization devices having very limited range of available tones. Driving printing devices is a typical application for digital halftoning algorithms. Many halftoning algorithms have been proposed in the past forty years. Nowadays, digital halftoning is considered as a mature topic. Nevertheless, as the technology of visualization devices undergoes continuous and steady progress, several algorithmic challenges in digital halftoning persist, as attest recent work in this field (Pang et al. 2008).

Most of the current halftoning algorithms produce printable images, adapted for a specific device. For example, error-diffusion algorithms perform well on ink-jet printers, where individual addressable dots (droplets of ink) are well-printed. However, the same algorithms perform rather poorly on laser or offset printers, where isolated dots may disappear completely (Kang 1999, Sharma 2002, Gupta & Bowen 2007). In addition, because of very non-linear behavior of electrostatic devices, inherent noise of error-diffusion may be greatly amplified by laser printers. Clustered-dot dithering produces excellent, stable and visually pleasant images on laser printers, but it performs rather poorly on ink-jet printers: the images appear too coarse and high-frequency low-contrast details may disappear. Offset printing or visualization of low-cost displays with limited number of available colors impose other constraints, and therefore require a very specific one-purpose halftoning algorithm. We concentrate in this work on dithering. The common practice when doing dithering is to use *threshold matrices*. Matrices are used essentially for commodity: matrices are simple to store and to manipulate. Figure 1 illustrates usage of threshold matrices for clustered-dot and dispersed-dot dithering. At the same time, rigid alignment of dither matrices can be rather harmful, as illustrated in Figure 6. Even tiny imperfections in spatial distribution of printed marks are immediately detected by our visual system. These artifacts are greatly amplified by non-linear behavior of conventional printing devices (namely, by dot gain in ink-jet or laser printing).

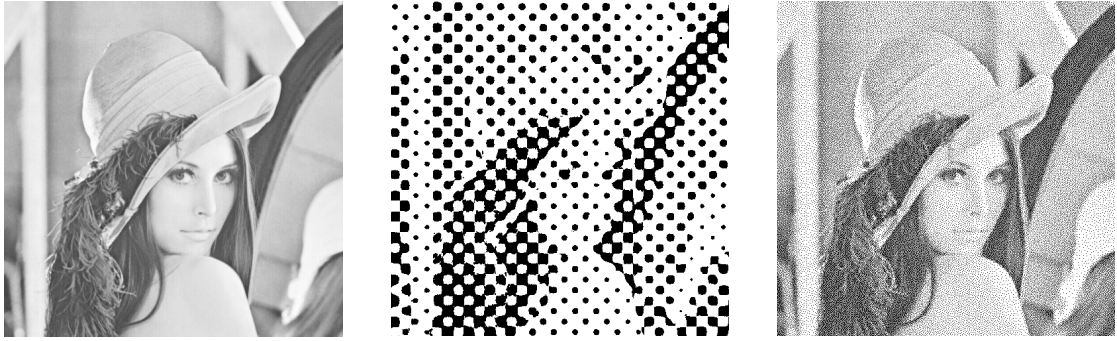

Figure 1. Illustration of halftoning using threshold matrices for both clustered (center) and dispersed (right) dot. The input image is shown on the left. A pre-computed matrix is tiled on the image plane, and input value are compared to each threshold matrix element to decide weather the output value is black or white

We propose to demonstrate that dither structures other than matrices can be advantageously used for the tasks where threshold matrices are usually employed. The structure we use is rectifiable polyominoes, a choice inspired by (Ostromoukhov 2007). Here the author convincingly has convincingly demonstrated the advantages of usage of non-periodic non-square structures instead of periodic square ones, and we take advantage of the same non-repetitive structure for our dithering method. A second important source of inspiration is the void-and-cluster method (Ulichney 1993), we use the simple and tractable optimization process describe by Ulichney. The main idea of this paper is rather simple: to replace square matrix of threshold values by a non-rectangle structure, which contains approximatively the same number of threshold values that as dither matrix. We propose to use rectifiable polyominoes: simple non-square figures which tile the Euclidean plane without gaps. The threshold values associated with the polyominoes can be optimized once forever, exactly as people do for building square blue-noise dither matrix (Kang 1999, Sharma 2002, Mitsa & Parker 1991, Lieberman & Allebach 2000). The optimized threshold values can be stored in lookup tables and inexpensively used during the image generation.

Figure 2a shows the basic principle of using of polyominoes in the context of digital halftoning. The image is entirely covered by a big polyomino. This polyomino is recursively subdivided until the pixel level is reached. At this time, appropriate threshold values are taken from the lookup tables where the optimized values of threshold are stored.

The rest of the paper is organized as follows. In Section 2 we give a short overview of the polyomino tiling, then, in Section 3, we describe the process of building and optimization of the polyomino-based threshold structure. We show some results in Section 4, and compare them to state-of-the-art methods. We draw conclusions in Section 5.

## 2. POLYOMINO TILING

### 2.1 Basic notions

We briefly recall here the definition of the polyomino tiling and its main properties, useful for our construction. The interested reader can find detailed descriptions in (Golomb 1996, Clarke 2006 Ostromoukhov 2007). *Polyomino* or *n-omino* is a plane topological disc, consisting of *n* edge-to-edge adjacent squares. Figure 2b-top shows a few simple polyominoes.

Polyominoes and their properties have been extensively studied in mathematics, and more precisely in combinatorial geometry (Grünbaum & Shephard 1986). A typical problem related to polyominoes can be formulated as follows: determine whether a given planar polygon can be filled, with no gaps, by a given set of polyominoes. For example, Figure 2b-bottom-left shows how the rectangle of dimensions 12x9 can be filled with 18 identical G-hexominoes.

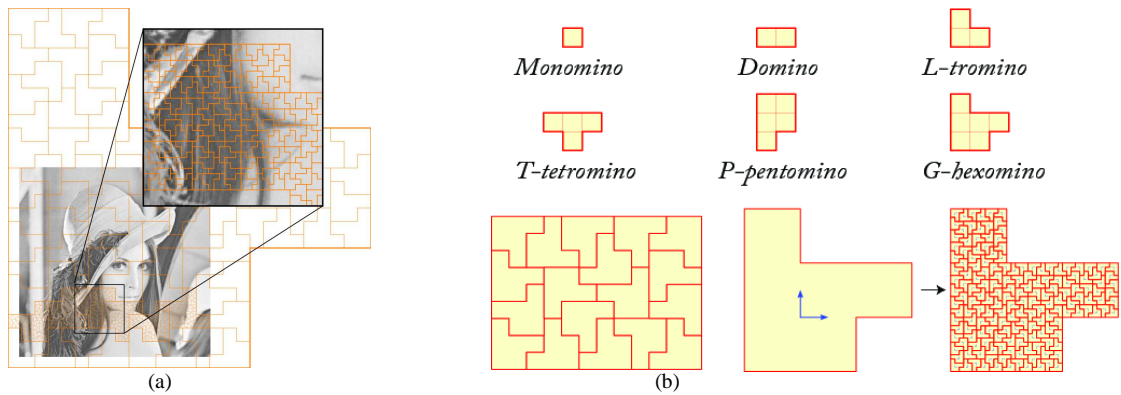

Figure 2: (a) The target image is entirely covered by a big polyomino, and then a deterministic process of subdivision is applied recursively until the pixel level is achieved. (b)-top Some polyomino shapes. (b)-bottom A 12x9 rectangle filled with 18 G-hexominoes. Production rule used for generating tiling with $9^2$-rep G-hexominoes

In this paper, we consider a special class of *rectifiable polyominoes*. A polyomino is said to be rectifiable if several copies of the polyomino form a rectangle. Rectifiable polyominoes can always be presented in terms of self-similar $\mathcal{L}^2$-rep constructions (also called *production rules*), where the larger version of the polyomino is built out $\mathcal{L}^2$ identical copies of the polyomino. Here, $\mathcal{L}$ is the linear scaling factor in the $\mathcal{L}^2$-rep construction, and $\mathcal{A} = \mathcal{L}^2$ is the area scaling factor. An example of a production rule for decomposition of the $9^2$-rep G-hexomino into $9^2$ identical G-hexominoes is shown in Figure 2b-bottom-right (in this case $\mathcal{L}=9$, and $\mathcal{A} = \mathcal{L}^2 = 9^2$).

Applying the production rules iteratively, and keeping the size of polyominoes constant, one can fill an arbitrarily large planar patch, and, at the limit, the entire plane (Grünbaum & Shephard 1986). For a given rectifiable polyomino, $\mathcal{L}$ is not unique: a variety of $\mathcal{L}^2$-rep constructions, for different linear scaling factors $\mathcal{L}$, can be found.

As it has been demonstrated in (Ostrmoukhov 2007), polyomino-based tilings show very distinct spectral property: the Fourier spectra of the tiles' centers is reasonably close to blue-noise spectra. This property is the consequence of the non-periodic, self-similar nature of $\mathcal{L}^2$-rep polyominoes. In fact, according to Statement 10.1.1 in (Grünbaum & Shephard 1986), if a monohedral $\mathcal{L}^2$-similarity tiling has a unique production rule, then such a tiling is not periodic. This facilitates building of isotropic distributions of points which exhibit blue-noise properties. In the context of digital halftoning, and more particularly, during the process of building the threshold structure, we shall also benefit from *a priori* good spectral properties of polyomino-based tilings which derive from the non-periodic self-similar nature of these tilings.

In summary, we take advantage of two important properties of rectifiable polyominoes. First, their construction is simple and deterministic; their geometrical properties can be exhaustively studied. Second, rectifiable polyominoes are fundamentally self-similar. Consequently, we can easily build a blue-noise threshold structure, which is inherently better than the square matrix-based Void-and-Cluster (Ulichney 1993) or Blue Noise Mask (Mitsa & Parker 1991) methods.

## 2.2 Structural Indices

Polyomino-based sampling systems (Grünbaum & Shephard 1986) used the key notion of *structural index*, which designates the local neighborhood of each tile. The idea is quite simple: polyominoes having identical neighborhoods, and consequently identical structural indices, will behave similarly in the process of optimization described in the following sections. Our goal is to identify geometrically identical configurations around each polyomino, and associate these identical configurations with the production rules.

Polyominoes are built of adjacent squares; the square's vertices form a square lattice. Let us mark each individual tile with letters around each lattice point as shown in Figure 3a. To facilitate the implementation, we decided to mark differently all four possible orientations of the polyominoes, together with their mirrored shapes. All we need is to walk through the tiling, identify all unique combinations of marks, and tabulate them.

Several properties related to structural indices can be proved. First, for a given production rule, the number of structural indices is finite. Second, subdivision of a polyomino having a certain structural index produces a unique combination of structural indices in the subdivided configuration. We tabulate the set of such configurations for all existing structural indices, and we call this table *structural indices production rules*. This table is used in every subdivision, according to the polyomino's attribute *structural indices*, which is an index to the structural indices production rules. Thus, starting with a polyomino having any structural index, we can deterministically define all structural indices of all polyominoes, after any number of subdivisions. Also, each pair of vertices of any polyomino having a given structural index can be uniquely identified by a set of label, as shown in Figure 3b. This facilitates the process of finding the optimal distribution of threshold values, as we shall explain in the next section.

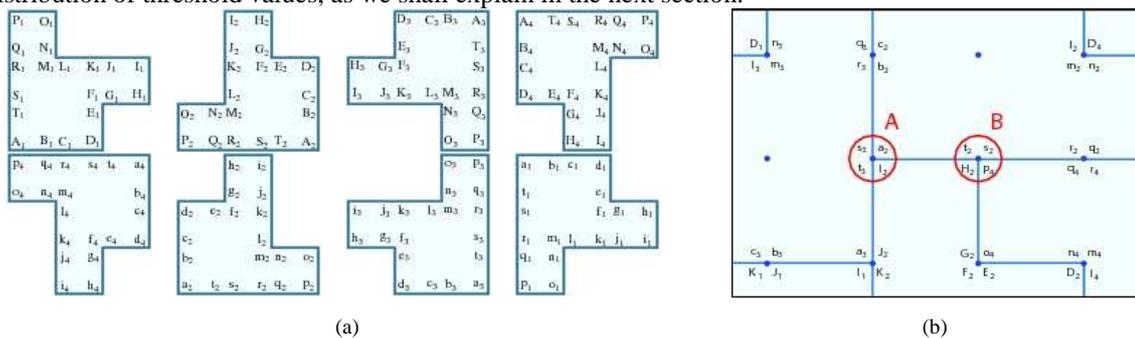

(a) (b)

Figure 3 (a) All these tiles have the same shape, but we consider them different by assigning them unique vertex identifiers. (b) Each border segment is uniquely determined by a combination of labels around each vertex. Here, the border segment (A, B) is defined by the labels around the vertices A and B: (t3,I2,a2,s3) and (H2,p4,s2,t2).

## 3. OPTIMIZATION OF THE THRESHOLD VALUES DISTRIBUTION

As we have already mentioned, our goal is to build a good distribution of thresholds values, somehow associated with polyominoes. Figure 4b illustrates the case of G-Hexominoes. Each G-Hexomino is formed of six squares; each of six squares contains $S^2$ distinct threshold values ($S = 8$ in Figure 4a). Every segment of polyomino's border has a specific ``border'' set of threshold values, which is at the same time part of the threshold structure (Figure 4a shows such 2-pixel-wide borders in red).

To avoid any ambiguity and duplication of the elements of the borders, we decided to put all border pixels left-and-upward with respect to the true polyomino's border (which indeed has no area, shown as blue lines in Figure 4b). The idea of dividing the whole set of threshold values into two subsets - polyomino's borders and polyomino's interior - is quite simple and efficient in the context of optimization: we first optimize the border threshold values, then we optimize the interior ones.

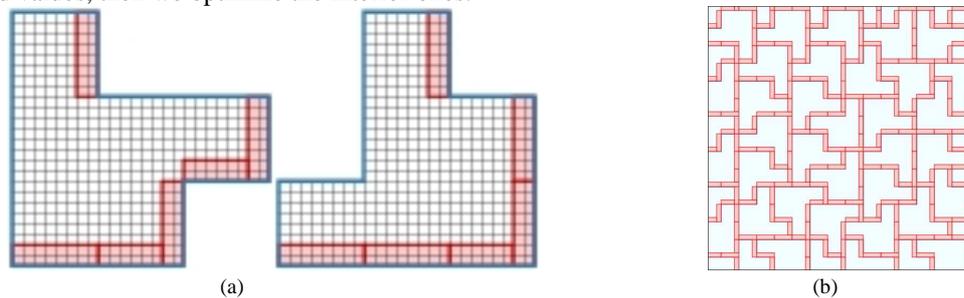

(a) (b)

Figure 4: (a) Association of a plurality of threshold values with the G-Hexominoes. Threshold values that belong to the border structure are highlighted in red. Please note that our border structure is not rotation-invariant. (b) The resulting distribution of borders associated with G-Hexominoes.

Each element or pixel which is part of the threshold structure associated with polyominoes contains a unique threshold value, used in the process of conventional dithering. Consequently, the process of finding the optimal distribution of the threshold values consists in the ranking of all elements within each polyomino. Polyominoes having different structural indices have different distributions of the rank values. During the optimization process, we need to find all rank values for all possible geometric combination that may occur in the tiling, and to store these ranking values in a lookup table, which can be consulted during the halftoning process.

Let us now describe in detail the optimization process. We start our optimization from a fixed intermediate density $d_0$ in [0, $S^2$], of black dots, separately processing first the border structure, and then the interior of each polyomino. Applying Lloyd relaxation, we obtain an excellent distribution of black pixels within the whole tiling. Then, we perform consecutive ranking, by either withdrawing pixels, for densities going toward 0, or by adding pixels, for densities going toward maximal density. The whole process can be subdivided into three stages: (a) initialization of the borders, (b) initial distribution of the interior pixels, and (c) consecutive ranking.

## 3.1 Initialization of the borders

The border structure is a well-identified subset of the tiling with G-Hexominoes, as shown in Figure 4b. We first optimize the distribution of black dots of the initial density $d_0$ within the border structure. This process is conceptually simple: first, we identify all existing border segments by their labels shown in Figure 3b. The list of all available border segments is tabulated. We process all border segments from this list in random order, one-by-one. For each border to be processed, we repeat the same algorithm, which consists in the following steps:
- Take a large area which contains a big number of tiles, e.g. the area shown in Figure 4b ;
- Put within the already-processed border segments the dots which are marked as ``fixed'';
- Put within the rest of the large area a certain number of dots, which corresponds (together with already-processed border segments) to the initial density $d_0$ ; mark these dots as ``floating'';
- Perform Lloyd relaxation;
- Select the dots that are in the current border segment being processed.

After this iterative relaxation process performed several times, we obtain a set of all border segments, with the distribution of dots being a good approximation of a blue-noise distribution and achieving the density $d_0$ .

## 3.2 Initial distribution of the interior pixels

Any polyomino which may occur in the tiling is surrounded by a set of border segments, well-identified by the combination of labels around each vertex of the polyomino, as explained in Section 2. Consequently, a polyomino with a unique combination of labels is necessarily surrounded by a well-determined set of border segments, for which the initial distribution of black dots has been already obtained, as explained in Section 3.1. These border segments have already a good distribution of dots which corresponds to initial density $d_0$. Consequently, the interior of each polyomino can be easily optimized under excellent ``border conditions''. The process follows these steps:
- Consider a large area which contains a big number of tiles;
- Add all border segments with their optimal set of dots, which are marked as ``fixed'';
- Put within each polyomino a certain number of dots, which corresponds (together with the border segments) to the initial density $d_0$ ; mark these dots as ``floating'';
- Perform Lloyd relaxation;
- Select the dots within the interior area and tabulate them, taking vertex labels as index.

Thus, we obtain an optimal distribution of dots of initial density $d_0$, within the whole tiling. Our construction is possible thanks to the following key observation: the border pixels are determinant during the relaxation process: these borders block the influence of the neighbors' interiors; the polyomino's interior *feels* only its border. Consequently, a unique set of vertex labels of each polyomino determines a unique optimal configuration of dots of initial density $d_0$ within the polyomino.

## 3.3 Consecutive ranking

Starting from the initial density $d_0$ of dots, we need to find the distribution of dots for all other densities. Our method follows the framework defined in (Ulichney 19993), but applied in a very specific context of tiling with polyominoes.

First, we perform ranking from $d_0$ down to 0. The process is as follows: First we choose a large area which contains a big number of tiles. For each tile, we consider all dots determined in Section 3.2. Then we iterate the following algorithm until all polyominoes have been processed:
- Pick a random polyomino from the tiling;
- Determine the pixel which, if withdrawn, forms the smallest ``hole", by performing Gaussian blur of appropriate kernel size on all possible candidates, and by choosing the best one;
- Mark the chosen pixel as being of rank $d_0$ -1.

Thus, we obtain a near-optimal distribution of dots of initial density $d_0$ -1, within the whole tiling. We repeat this operation iteratively until the rank 0. Next, we perform ranking from $d_0$ up to $S^2$. This process is very similar to that used for down-ranking, with one noticeable difference: instead of withdrawing pixels, we set them among available positions within the polyomino. The criterion for selecting is the smallest ``cluster" which is formed.

## 4. RESULTS

The threshold structure obtained during the optimization process described in the previous section possesses many important properties: it shows nearly-optimal blue-noise distribution for all gray levels, as illustrated in Figure 5. A side-by-side comparison with the void-and-cluster method is quite advantageous for us. As expected, our method hides alignments artifact present in traditional matrix-based dithering due to the finite size of the threshold matrices. Please notice that we used for our comparison, in Figure 6, the threshold structures of approximatively equal size.

The proposed method can be used in any context where the blue-noise threshold matrices are traditionally used for a practical application: in clustered-dot or dispersed-dot dithering, error diffusion with threshold modulation, etc. We provide here only one simple illustration, see Figure 7.

Our implementation is computationally efficient. The optimization process is expensive, but it is performed only once. The runtime application of the tiling and selecting appropriate threshold values is truly inexpensive. In fact, all production rules used for polyomino generation are tabulated and do not require any complicated calculations; the set of threshold valued associated with the polyominoes of different combination of labels are equally tabulated. All label calculations are performed during the pre-process; the production rules and the threshold value indexing is performed according to these pre-computed and tabulated indices (this principle has been introduced in Ostromoukhov 2007).

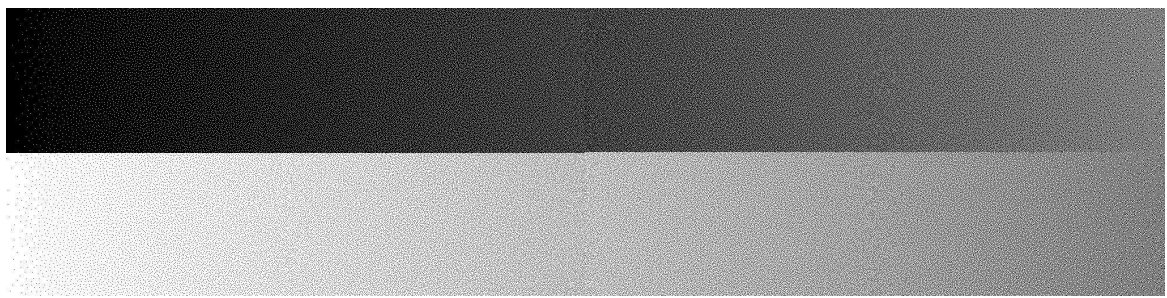

Figure 5: A set of dither ramps from 0 to 1 dithered with our method. We used a base matrix size $S = 8$ and initial density $d_0 = 1 / S$

## 5. CONCLUSION

In this paper, we have demonstrated that polyominoes, and more particularly G-Hexominoes, can be advantageously used for building threshold structures used in various halftoning algorithms. Compared to the matrices produced with previous state-of-the-art algorithms, our threshold structure of comparable size shows superior quality. Thanks to inherently non-periodic nature of the polyomino tiling, border-alignment artifacts which are slightly visible in concurrent methods becomes completely invisible in our case. Our method is in addition computationally inexpensive and relatively easy to implement.

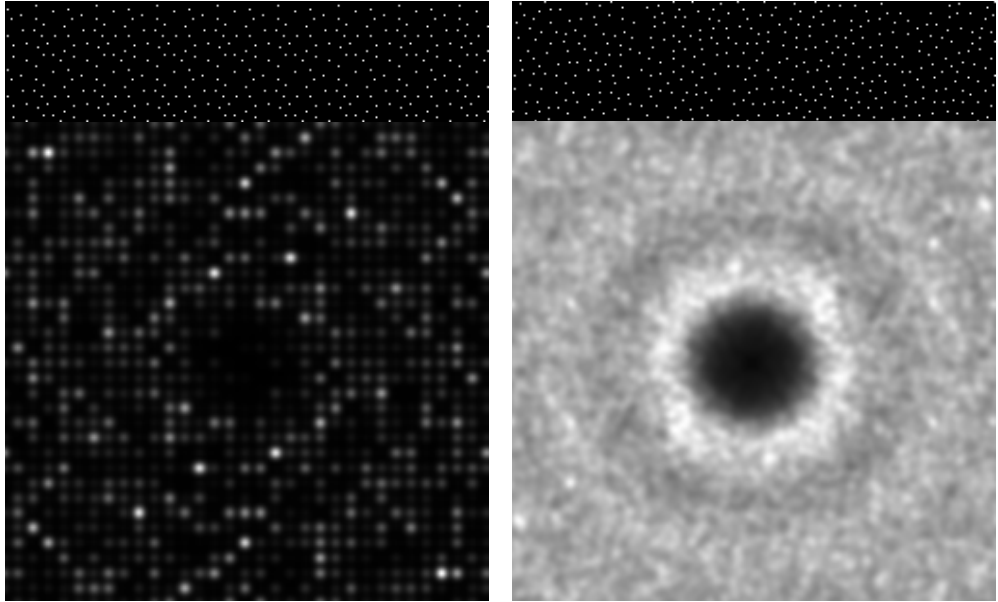

Figure 6: Top line shows a sample of level 6/256, bottom is the corresponding power spectrum (sum of Fourier spectra of 10 random sample of size 256x256, the result is blurred to emphasis high energy peaks). Left is void-and-cluster, right is our method

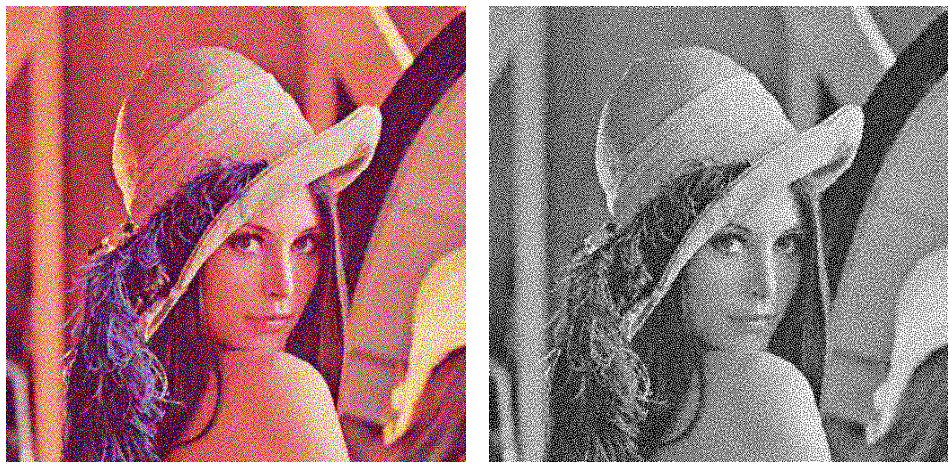

Figure 7: Two sample images (color and black/white) showing applicability of our method for dispersed-dot dithering.

# REFERENCES

A. L. Clarke, 2006 The Poly Pages http://www.recmath.com/PolyPages.

S. W. Golomb, 1996 *Polyominoes: Puzzles, Patterns, Problems, and Packings.* Princeton University Press.

B. Grünbaum and G. Shephard, 1986. *Tilings and Patterns*. W.H. Freeman.

M. R. Gupta and J. Bowen, 2007. Ranked dither for robust color printing. *Proceedings of SPIE*. San Jose CA, USA.

H. R. Kang, 1999 *Digital Color Halftoning*. SPIE, Bellingham, WA, USA.

D. J. Lieberman and J. P. Allebach, 2000. A dual interpretation for direct binary search and its implications for tone reproduction and texture quality. *In IEEE Transactions on Image Processing*, Vol. 9, No 11, pp 1950--1963.

T. Mitsa and K. J. Parker, 1991 Digital halftoning using a blue noise mask. *ICASSP~91: 1991 International Conference on Acoustics, Speech, and Signal Processing*, Vol. 2, pp 2809-2812.

V. Ostromoukhov, 2007. Sampling with polyominoes. *ACM Transactions on Graphics (SIGGRAPPH)*, Vol. 26, No. 3 pp 78:1-78:6.

W.-M. Pang, Y. Qu, T.-T. Wong, D. Cohen-Or, and P.-A. Heng. Structure-aware halftoning. *ACM Transactions on Graphics ( SIGGRAPH).* Vol. 27, No 3.

G. Sharma, 2002. *Digital Color Imaging Handbook*. CRC Press, Inc., Boca Raton, FL, USA.

R. Ulichney, 1993. The void-and-cluster method for generating dither arrays. *Proceedings of SPIE.* pp 332-343.